\documentclass[showpacs,preprintnumbers,amsmath,amssymb,floatfix]{revtex4}

\usepackage{bm}
\usepackage{epsfig}
\usepackage{amssymb}
\usepackage{amsmath}
\usepackage{calrsfs}
\usepackage{multirow}
\usepackage{rotating}
\usepackage{xcolor}
\begin{document}

\setlength{\arraycolsep}{0.5mm}

\preprint{DESY 17-020, NSF-KITP-17-018\hspace{10.5cm}
ISSN 0418--9833}
\preprint{February 2017\hspace{15.6cm}}

\title{$\alpha_s$ from hadron multiplicities via SUSY-like relation between
anomalous dimensions}

\author{Bernd~A.~Kniehl}
\thanks{Permanent address: 
{II.} Institut f\"ur Theoretische Physik,
Universit\"at Hamburg, Luruper Chaussee 149, 22761 Hamburg, Germany.}
\affiliation{Kavli Institute for Theoretical Physics, University of California,
Santa Barbara, CA~93106-4030, USA}

\author{Anatoly~V.~Kotikov}
\affiliation{Bogoliubov Laboratory of Theoretical Physics,
Joint Institute for Nuclear Research, 141980 Dubna, Russia}

\begin{abstract}
We recover in QCD an amazingly simple relationship between the anomalous
dimensions, resummed through next-to-next-to-leading-logarithmic order, in the
Dokshitzer-Gribov-Lipatov-Altarelli-Parisi evolution equations for the
first Mellin moments $D_{q,g}(\mu^2)$ of the quark and gluon fragmentation
functions, which correspond to the average hadron multiplicities in jets
initiated by quarks and gluons, respectively.
This relationship, which is independent of the number of quark flavors,
dramatically improves previous treatments by allowing for an exact solution of
the evolution equations.
So far, such relationships have only been known from supersymmetric QCD, where
$C_F/C_A=1$.
This also allows us to extend our knowledge of the ratio
$D_g^-(\mu^2)/D_q^-(\mu^2)$ of the minus components by one order in
$\sqrt{\alpha_s}$. 
Exploiting available next-to-next-to-next-to-leading-order information on
the ratio $D_g^+(\mu^2)/D_q^+(\mu^2)$ of the dominant plus components, we fit the
world data of $D_{q,g}(\mu^2)$ for charged hadrons measured in $e^+e^-$
annihilation to obtain
$\alpha_s^{(5)}(M_Z)=0.1205\genfrac{}{}{0pt}{}{+0.016}{-0.0020}$. 
\end{abstract}


\maketitle

In the parton model of QCD \cite{Bjorken:1969ja}, the inclusive
production of single hadrons involves the notion of fragmentation functions
$D_a(x,\mu^2)$, where $\mu$ is the factorization scale.
At leading order (LO), their values correspond to the probability for a parton
$a=q,g$ produced at short distance $\hbar c/\sqrt{\mu^2}$ to produce a jet that
contains a hadron carrying the fraction $x$ of the momentum of parton $a$.
Owing to the factorization theorem, the $D_a(x,\mu^2)$ functions are universal
in the sense that they do not depend on the process by which parton $a$ is
produced.
By local parton-hadron duality \cite{Azimov:1984np}, there should be a local
correspondence between parton and hadron distributions in hard-scattering
processes.
Yet, $D_a(x,\mu^2)$ are genuinely nonperturbative, which implies
that their $x$ dependences at some scale $\mu_0$ cannot be calculated from the
QCD Lagrangian using perturbation theory, but need to be determined by fitting
experimental data.
However, once $D_a(x,\mu_0^2)$ are assumed to be known, their $\mu^2$
dependences
are governed by the timelike Dokshitzer-Gribov-Lipatov-Altarelli-Parisi (DGLAP)
evolution equations \cite{Gribov:1972ri,Dokshitzer:1977sg}.
The anomalous dimensions therein, the $a\to b$ splitting functions $P_{ba}(x)$,
are known at next-to-next-to-leading order \cite{Almasy:2011eq}.
The scaling violations, {\it i.e.}, the $\mu^2$ dependences, of $D_a(x,\mu^2)$
may be exploited in global data fits to extract the strong-coupling constant
$\alpha_s=g_s^2/(4\pi)$, leading to very competitive results
\cite{Kniehl:2000fe} as for the world average \cite{Olive:2016xmw}.

The DGLAP equations are conveniently solved in Mellin space, where
$D_a(N,\mu^2)=\int dx\,x^{N-1}D_a(x,\mu^2)$ with $N=1,2,\ldots$ and similarly for
$P_{ba}(x)$, because convolutions are converted to products.
We have
\begin{equation}
\frac{\mu^2d}{d\mu^2}
\left(\begin{array}{l} D_s(N,\mu^2) \\ D_g(N,\mu^2) \end{array}\right)
=\left(\begin{array}{ll} P_{qq}(N) & P_{gq}(N) \\
P_{qg}(N) & P_{gg}(N) \end{array}\right)
\left(\begin{array}{l} D_s(N,\mu^2) \\ D_g(N,\mu^2) \end{array}\right),
\label{apR}
\end{equation}
where $D_s=(1/2n_f)\sum_{q=1}^{n_f}(D_q+D_{\bar{q}})$, with $n_f$ being the
number of active quark flavors, is the quark singlet component.
The quark non-singlet component, which is irrelevant for the following, obeys
a decoupled DGLAP equation.
After solving the DGLAP equations in Mellin space, one returns to $x$ space via
the inverse Mellin transform, analytically continuing $N$ to complex values.

The first Mellin moment $D_a(\mu^2)\equiv D_a(1,\mu^2)$ is of particular
interest in its own right because, up to corrections of orders beyond our
consideration here, it corresponds to the average hadron multiplicity
$\langle n_h\rangle_a$ of jets initiated by parton $a$.
There exists a wealth of experimental data on $\langle n_h\rangle_q$,
$\langle n_h\rangle_g$, and their ratio
$r=\langle n_h\rangle_g/\langle n_h\rangle_q$ for charged hadrons $h$ taken in
$e^+e^-$ annihilation at various center-of-mass energies $\sqrt{s}$, ranging
from 10 to 209~GeV (for a comprehensive compilation of experimental
publications, see Ref.~\cite{Bolzoni:2013rsa}), which allows for a
high-precision determination of $\alpha_s$
\cite{Bolzoni:2013rsa,Perez-Ramos:2013eba}.
In fact, besides $\alpha_s$ and ignoring power corrections for the time being,
there are just two more fit parameters, $D_q(\mu_0^2)$ and $D_g(\mu_0^2)$ at
some reference scale $\mu_0$, which have a very clear and simple physical
interpretation, while no input from external sources, {\it e.g.}, parton
density functions, is required.
This provides a strong motivation for us to deepen our theoretical
understanding of $D_a$ within the QCD formalism as much as possible, which is
actually limiting the error in the value of $\alpha_s$ thus extracted.
The study of $D_a$ is a topic of old vintage; the LO value of $r$,
$C^{-1}=C_A/C_F$ with color factors $C_F=4/3$ and $C_A=3$, was found four
decades ago \cite{Brodsky:1976mg}.
Subsequent analyses \cite{Perez-Ramos:2013eba,Malaza:1985jd} were performed
using the generating-functional approach in the modified leading-logarithmic
approximation (MLLA) \cite{Dokshitzer:1991wu}.

The description of the $\mu^2$ dependences of $D_a$ at fixed order in
perturbation theory are spoiled by the fact that $P_{ba}\equiv P_{ba}(1)$ are
ill defined and require resummation, which was performed for the leading
logarithms (LL) \cite{Mueller:1981ex}, the next-to-leading logarithms (NLL)
\cite{Vogt:2011jv}, and the next-to-next-to-leading logarithms (NNLL)
\cite{Kom:2012hd}.
In Ref.~\cite{Bolzoni:2013rsa}, Eq.~(\ref{apR}) is first
diagonalized for arbitrary value of $N$ at LO, and then the NNLL resummation is
incorporated.
Unfortunately, this two-step procedure, which has been standard practice in the
literature so far \cite{Buras:1979yt,Ellis:1993rb}, fails to fully exploit the
available knowledge on the higher-order corrections and yields an
approximation, the uncertainty of which is difficult to estimate reliably.

In this Letter, we expose a relationship between the NNLL-resummed expressions
for $P_{ba}$, which has gone unnoticed so far.
Its existence in QCD is quite remarkable and interesting in its own right,
because a similar relationship is familiar from supersymmetric (SUSY) QCD,
where $C=1$
\cite{Dokshitzer:1977sg,Dokshitzer:1991wu,Kom:2012hd,Kounnas:1982de}.
Owing to this new relationship, the DGLAP equations may be solved exactly,
which greatly consolidates the theoretical foundation for the determination of
$\alpha_s$ and thus reduces its theoretical uncertainty.

Our starting point is Eq.~(\ref{apR}) for $N=1$ with NNLL resummation.
We have \cite{Kom:2012hd}
\begin{eqnarray}
P_{aa}&=&\gamma_0(\delta_{ag} + K_{a}^{(1)} \gamma_0 
+ K_{a}^{(2)}  \gamma_0^2) + \mathcal{O}(\gamma_0^4)\qquad(a=q,g),
\nonumber\\
P_{gq}&=& C (P_{gg} +A)
+ \mathcal{O}(\gamma_0^4),
\nonumber\\
P_{qg} &=& 
C^{-1} (P_{qq} +A) 
+ \mathcal{O}(\gamma_0^4),
\label{NNLL}
\end{eqnarray}
where $\gamma_0=\sqrt{2C_Aa_s}$, with $a_s=\alpha_s/(4\pi)$ being the couplant,
$\delta_{ab}$ is the Kronecker symbol, and
\begin{eqnarray}
K_{q}^{(1)} &=& \frac{2}{3} C\varphi,\quad
K_{q}^{(2)} = -\frac{1}{6} C\varphi [17-2\varphi(1-2C)],
\nonumber \\
K_{g}^{(1)} &=& -\frac{1}{12}[11 +2\varphi (1+6C)],\quad 
K_{g}^{(2)} = \frac{1193}{288} -2\zeta(2)
- \frac{5\varphi}{72}(7-38C)+\frac{\varphi^2}{72}(1-2C)(1-18C),
\nonumber \\
A &=& K_{q}^{(1)}\gamma_0^2,\quad
\varphi=\frac{2n_fT_R}{C_A},\quad
T_R=\frac{1}{2}.
\label{nllfirstA}
\end{eqnarray}
Eq.~(\ref{NNLL}) is written in a form that allows us to glean a novel
relationship:
\begin{equation}
C^{-1}P_{gq}-P_{gg}=CP_{qg}-P_{qq},
\label{Basic}
\end{equation}
which is independent of $n_f$.
Eq.~(\ref{Basic}) generalizes the case of SUSY QCD
\cite{Dokshitzer:1977sg,Dokshitzer:1991wu,Kom:2012hd,Kounnas:1982de} from
$C=1$ to $C=9/4$.
The corresponding relation in $\mathcal{N}=1$ SUSY
\cite{Dokshitzer:1977sg} is known to be violated beyond LO 
\cite{Almasy:2011eq}.
It will be interesting to see if Eq.~(\ref{Basic}) also holds beyond
$\mathcal{O}(\gamma_0^3)$.

We now solve Eq.~(\ref{apR}) exactly by exploiting Eq.~(\ref{Basic}).
To this end, we diagonalize the NNLL DGLAP evolution kernel as
\begin{equation}
U^{-1}\left(\begin{array}{ll}
P_{qq} & P_{gq} \\ P_{qg} & P_{gg}
\end{array}\right)U
=\left(\begin{array}{ll}
P_{--} & 0 \\ 0 & P_{++}
\end{array}\right),
\end{equation}
by means of the matrices \cite{Buras:1979yt}
\begin{equation}
U=\left(\begin{array}{ll}
1 & -1 \\ \frac{1-\alpha}{\varepsilon} & \frac{\alpha}{\varepsilon}
\end{array}\right),\qquad
U^{-1}=\left(\begin{array}{ll} 
\alpha & \varepsilon \\ \alpha-1 & \varepsilon
\end{array}\right),
\label{matrix}
\end{equation}
where 
\begin{eqnarray}
\alpha&=&\frac{P_{qq}-P_{++}}{P_{--}-P_{++}},
\qquad
\varepsilon=\frac{P_{gq}}{P_{--}-P_{++}},
\\
P_{\pm\pm}&=&\frac{1}{2}\left[P_{qq}+P_{gg}\pm
\sqrt{(P_{qq}-P_{gg})^2+4P_{qg}P_{gq}}\right].\qquad
\label{Ppm}
\end{eqnarray}
Eq.~(\ref{apR}) thus assumes the form
\begin{equation}
\frac{\mu^2d}{d\mu^2}\left(\begin{array}{l} D_- \\ D_+ \end{array}\right)
=\left[
\left(\begin{array}{ll} P_{--} & 0 \\ 0 & P_{++}\end{array}\right)
-U^{-1}\frac{\mu^2d}{d\mu^2}U\right]
\left(\begin{array}{l} D_- \\ D_+ \end{array}\right),
\label{ap2a}
\end{equation}
where the second term contained within the square brackets stems from the
commutator of $\mu^2d/d\mu^2$ and $U$, and
\begin{equation}
\left(\begin{array}{l} D_- \\ D_+ \end{array}\right)
=U^{-1}\left(\begin{array}{l} D_s \\ D_g \end{array}\right)
=\left(\begin{array}{l}(\alpha-1)D_s+\varepsilon D_g \\
\alpha D_s+\varepsilon D_g \end{array}\right).
\label{ap1.2}
\end{equation}
Owing to Eq.~(\ref{Basic}), the square root in
Eq.~(\ref{Ppm}) disappears, and we have
\begin{eqnarray}
P_{--}&=&-A,\qquad
P_{++}=P_{qq}+P_{gg}+A,
\label{P+-}
\\
\alpha&=&\frac{P_{gg}+A}{P_{qq}+P_{gg}+2A},\qquad
\varepsilon = -C \alpha \, .
\label{alpha1}
\end{eqnarray}
Inserting the second equality of Eq.~(\ref{alpha1}) in Eq.~(\ref{matrix}), we
have
\begin{equation}
U^{-1}\frac{\mu^2d}{d\mu^2}U=-\frac{1}{\alpha}\,\frac{\mu^2d}{d\mu^2}\alpha
\left(\begin{array}{ll} 1 & 0 \\ 1 & 0\end{array}\right).
\label{AddTerm}
\end{equation}
Using the QCD $\beta$ function,
\begin{equation}
\frac{\mu^2d}{d\mu^2}a_s=\beta(a_s)=-\beta_0a_s^2-\beta_1a_s^3+\mathcal{O}(a_s^4),
\label{eq:beta}
\end{equation}
with one- and two-loop coefficients
\begin{equation}
\beta_0=\frac{C_A}{3}(11-2\varphi),\quad
\beta_1=\frac{2C_A^2}{3}[17-\varphi(5+3C)],
\end{equation}
we may convert the differential operator as
\begin{equation}
\frac{\mu^2d}{d\mu^2}=\frac{C_A}{\gamma_0}
\beta\left(\frac{\gamma_0^2}{2C_A}\right)\frac{d}{d\gamma_0}.
\label{eq:con}
\end{equation}
Inserting Eqs.~(\ref{NNLL}) and (\ref{nllfirstA}) in the first equality of
Eq.~(\ref{alpha1}), we have
$\alpha=1-4C\varphi\gamma_0/3+\mathcal{O}(\gamma_0^2)$, so that
\begin{equation}
\frac{1}{\alpha}\,\frac{\mu^2d}{d\mu^2}\alpha
=\frac{C\varphi\beta_0}{3C_A}\gamma_0^3+\mathcal{O}(\gamma_0^4).
\label{eq:da}
\end{equation}
Inserting Eqs.~(\ref{P+-}), (\ref{AddTerm}), and (\ref{eq:da}) in
Eq.~(\ref{ap2a}), we may cast Eq.~(\ref{apR}) in its final form,
\begin{equation}
\frac{\mu^2d}{d\mu^2}
\left(\begin{array}{l} D_- \\ D_+ \end{array}\right)
= \left(\begin{array}{ll} \frac{C\varphi\beta_0}{3C_A}\gamma_0^3-A & 0 \\ 
\frac{C\varphi\beta_0}{3C_A}\gamma_0^3 & P_{gg}+P_{qq} + A \end{array}\right) 
\left(\begin{array}{l} D_- \\ D_+ \end{array}\right).\quad
\label{ap2b}
\end{equation}
The initial conditions are given by Eq.~(\ref{ap1.2}) for $\mu=\mu_0$ in terms
of the three constants $\alpha_s(\mu_0^2)$, $D_s(\mu_0^2)$, and $D_g(\mu_0^2)$.

The solution of Eq.~(\ref{ap2b}) is greatly facilitated by the fact that one
entry of the matrix on its right-hand side is zero.
We may thus obtain $D_-$ as the general solution of a homogeneous differential
equation, 
\begin{eqnarray}
\frac{D_-(\mu^2)}{D_-(\mu_0^2)} &=&
\exp\!{\left[\int_{\mu_0^2}^{\mu^2}\!\!\!\frac{d\bar{\mu}^2}{\bar{\mu}^2}
    \!\left(\frac{C\varphi\beta_0}{3C_A}\gamma_0^3-A\!\right)\!\right]}
\nonumber\\
&=& \frac{T_-(\gamma_0(\mu^2))}{T_-(\gamma_0(\mu_0^2))}, 
\label{gensol.-}
\end{eqnarray}
where, with the help of Eq.~(\ref{eq:con}),
\begin{eqnarray}
T_-(\gamma_0)&=&\exp{\left[\frac{4C\varphi}{3}\int d\gamma_0
\left(\frac{2C_A}{\beta_0\gamma_0 }-1\right)\right]} 
\nonumber\\
&=&\gamma_0^{d_-}\exp{\left(-\frac{4}{3}C\varphi \gamma_0\right)},
\label{T-}
\label{eq:t-}
\end{eqnarray}
with $d_-=8C_AC\varphi/(3\beta_0)$.
The small-$x$ correction $\propto\gamma_0$ in Eq.~(\ref{eq:t-}) originates from
the extra term in Eq.~(\ref{ap2a}) and represents a novel feature of our
approach.
In Ref.~\cite{Bolzoni:2013rsa} and analogous analyses for
parton distribution functions \cite{Kotikov:1998qt}, the minus
components do not participate in the resummation.

We are then left with an inhomogeneous differential equation for $D_+$.
The general solution $\tilde{D}_+$ of its homogeneous part reads
\begin{eqnarray}
\frac{\tilde{D}_+(\mu^2)}{\tilde{D}_+(\mu_0^2)}&=&
\exp{\left[\int_{\mu_0^2}^{\mu^2}\frac{d\bar{\mu}^2}{\bar{\mu}^2}
\gamma_0\left(1+ K_+^{(1)}\gamma_0+K_+^{(2)}\gamma_0^2\right)\right]}
\nonumber\\
&=&\frac{T_+(\gamma_0(\mu^2))}{T_+(\gamma_0(\mu_0^2))},
\label{gensol.+}
\end{eqnarray}
where
\begin{eqnarray}
K_+^{(1)}&=&2K_{q}^{(1)}+K_{g}^{(1)} =
-\frac{1}{12}[11 +2\varphi (1-2C)],
\nonumber\\
K_+^{(2)}&=&K_{q}^{(2)}+K_{g}^{(2)} 
= \frac{1193}{288} -2\zeta(2) - \frac{7\varphi}{72}(5+2C) 
+\frac{\varphi^2}{72}(1-2C)(1+6C),
\nonumber\\
T_+(\gamma_0)&=&\exp{\left[-\frac{4C_A}{\beta_0}
\int\frac{d\gamma_0}{\gamma_0^2}\,
\frac{1+ K^{(1)}_{+}\gamma_0 +  K^{(2)}_{+}\gamma_0^2}{1+b_1\gamma_0^2}
\right]} 
\nonumber\\
&=&\gamma_0^{d_+}\exp{\left[\frac{4C_A}{\beta_0\gamma_0}
-\frac{4C_A}{\beta_0} \left(K_+^{(2)}-b_1\right)\gamma_0\right]},
\label{T+}
\end{eqnarray}
with $d_+=-4C_AK_+^{(1)}/\beta_0$ and $b_1=\beta_1/(2C_A \beta_0)$.
Adding to $\tilde{D}_+$ a special solution of the inhomogeneous differential
equation for $D_+$, we find its general solution to be 
\begin{equation}
D_+(\mu^2)=\left[\frac{D_+(\mu_0^2)}{T_+(\gamma_0(\mu_0^2))}
-\frac{4}{3}C\varphi\frac{D_-(\mu_0^2)}{T_-(\gamma_0(\mu_0^2))}
\int_{\gamma_0(\mu_0^2)}^{\gamma_0(\mu^2)}
\frac{d\gamma_0}{1+b_1\gamma_0^2}\,\frac{T_-(\gamma_0)}{T_+(\gamma_0)}\right]
T_+(\gamma_0(\mu^2)).
\label{gensol.+a}
\end{equation}
The final expressions for $D_-$ and $D_+$ in Eqs.~(\ref{gensol.-}) and
(\ref{gensol.+a}), respectively, are fully renormalization group improved
because all $\mu$ dependence resides in $\gamma_0$.

The NLL approximation is recovered by omitting the exponential factor
multiplying $\gamma_0^{d_-}$ in Eq.~(\ref{T-}), putting $K_+^{(2)}=b_1=0$ in
Eq.~(\ref{T+}), and omitting the second term within the square brackets in
Eq.~(\ref{gensol.+a}).
The LL approximation then emerges from the NLL one by also putting $d_-=d_+=0$
in Eqs.~(\ref{T-}) and (\ref{T+}), respectively.
Hence follows the large-$\mu^2$ asymptotic behavior
$D_-/D_+\propto\exp\{-[(8C_A/\beta_0)\ln(\mu^2/\Lambda^2)]^{1/2}\}$, where
$\Lambda$ is the asymptotic scale parameter, which implies a strong fading of
$D_-$.

Using Eqs.~(\ref{matrix}) and (\ref{ap1.2}), we now return to the parton basis,
where it is useful to decompose $D_a=D_a^++D_a^-$ into the large and small
components $D_a^\pm$ proportional to $D_\pm$, respectively.
Defining $r_\pm=D_g^\pm/D_s^\pm$ and using Eqs.~(\ref{NNLL}), (\ref{nllfirstA}),
and (\ref{alpha1}), we then have $D_s^\pm=\mp D_\pm$ and
\begin{eqnarray}
r_+&=&-\frac{\alpha}{\epsilon}=\frac{1}{C}+\mathcal{O}(\gamma_0^2),
\label{eq:rpm}\\
r_-&=&\frac{1-\alpha}{\epsilon}
=-\frac{4}{3}\varphi\gamma_0
+\frac{\varphi}{18}[29-2\varphi(5-2C)]\gamma_0^2
+\mathcal{O}(\gamma_0^3).
\nonumber
\end{eqnarray}
Recalling that $\langle n_h\rangle_q=D_s$ and $\langle n_h\rangle_g=D_g$, we
thus have
\begin{equation}
r=\frac{r_++r_-D_s^-/D_s^+}{1+D_s^-/D_s^+}.
\label{eq:r}
\end{equation}
Eq.~(\ref{eq:rpm}) differs from Eqs.~(53) and (54) in
Ref.~\cite{Bolzoni:2013rsa},
\begin{eqnarray}
\bar{r}_+&=&\frac{1}{C}\left\{1-\frac{\gamma_0}{3}[1+\phi(1-2C)]
+\mathcal{O}(\gamma_0^2)\right\},
\nonumber\\
\bar{r}_-&=&-\frac{2}{3}\phi\gamma_0+\mathcal{O}(\gamma_0^2).
\label{eq:npb}
\end{eqnarray}
On the other hand, $\bar{r}_+$ in Eq.~(\ref{eq:npb}) agrees with the result for
$r$ obtained in Ref.~\cite{Mueller:1983cq} in the approximation of putting
$D_a^-=0$ and extended to through $\mathcal{O}(\gamma_0^3)$ in
Refs.~\cite{Dremin:1993vq,Capella:1999ms}, which is in line with the reasoning
in Chapter~7 of Ref.~\cite{Dokshitzer:1991wu}.
By the same token, we may accommodate the higher-order corrections
\cite{Dremin:1993vq,Capella:1999ms} by including within the curly brackets in
Eq.~(\ref{eq:npb}) the terms 
$\bar{c}_2\gamma_0^2+\bar{c}_3\gamma_0^3$, where
\begin{eqnarray}
\bar{c}_2&=&\frac{179}{72}-\frac{20}{9}\zeta(2)-\frac{355}{1944}n_f
+\frac{43}{26244}n_f^2,
\nonumber\\
\bar{c}_3&=&-\frac{5213}{1152}-\frac{8}{3}\zeta(2)+\frac{40}{9}\zeta(3)
+\left(-\frac{9761}{31104}+\frac{14}{27}\zeta(2)\right)
n_f+\frac{15595}{314928}n_f^2
-\frac{4799}{17006112}n_f^3.
\label{eq:co}
\end{eqnarray}
For $n_f=5$,
\begin{equation}
\bar{r}_+=2.250-0.889\,\gamma_0-4.593\,\gamma_0^2+0.740\,\gamma_0^3
+\mathcal{O}(\gamma_0^4).
\label{eq:rp5}
\end{equation}
The difference between $r_\pm$ and $\bar{r}_\pm$ is an artifact of the
different
diagonalization procedures adopted here and in Ref.~\cite{Bolzoni:2013rsa}.
In fact, taking the limit $N\to1$ in $D_a(N,\mu^2)$ and diagonalizing the DGLAP
equations are noncommuting operations.
Consequently, our components $D_a$ differ from those in
Ref.~\cite{Bolzoni:2013rsa}, $\overline{D}_a$ with
$\bar{r}_\pm=\overline{D}_g^\pm/\overline{D}_s^\pm$, by terms of
$\mathcal{O}(\gamma_0)$.
Specifically, we have
\begin{equation}
D_a^\pm=\sum_{b=s,g}M_{ab}\overline{D}_b^\pm,
\label{transform}
\end{equation}
where
\begin{eqnarray}
M_{ss}&=&1-\frac{4}{3}C\varphi\gamma_0,\qquad
M_{sg}= - \frac{C}{3} \gamma_0 [1+\varphi(1-6C)],\nonumber\\
M_{gs}&=&-\frac{2}{3}\varphi\gamma_0,\qquad
M_{gg}=1+\frac{2}{3}C\varphi\gamma_0.
\end{eqnarray}
In fact, this transformation converts $\bar{r}_\pm$ into $r_\pm$ and, by
exploiting Eq.~(\ref{eq:co}), allows us to extend our result for $r_+$ through
$\mathcal{O}(\gamma_0^3)$; the counterpart of Eq.~(\ref{eq:rp5}) reads 
\begin{equation}
r_+=2.250-4.505\,\gamma_0^2-0.586\,
\gamma_0^3 +\mathcal{O}(\gamma_0^4).
\label{eq:rpcap}
\end{equation}
Note that our advanced procedure to solve Eq.~(\ref{apR}) allows us to
determine $r_-$ through $\mathcal{O}(\gamma_0^2)$, while $\bar{r}_-$ from
Ref.~\cite{Bolzoni:2013rsa} is limited to $\mathcal{O}(\gamma_0)$.
We denote the approximation of using Eq.~(\ref{eq:rpcap}) on top of
Eqs.~(\ref{eq:rpm}) and (\ref{eq:r}) as NNNLO${}_\mathrm{approx}$+NNLL.

Power-like corrections were found to be indispensable for a realistic
description of the experimental data of 
$\langle n_h\rangle_q$, $\langle n_h\rangle_g$, and $r$
\cite{Capella:1999ms,Dokshitzer:1992iy}.
Following Refs.~\cite{Capella:1999ms,Dokshitzer:1992iy}, we include them by
multiplying $r_+$ in Eq.~(\ref{eq:rpcap}) with the factor
\begin{equation}
1+(1+\frac{n_f}{27})\frac{\mu_{\mathrm{cr}}}{\mu}\gamma_0,
\label{eq:power}
\end{equation}
where $\mu_\mathrm{cr}$ is a critical scale parameter to be fitted.
In the MLLA approach, $\mu_\mathrm{cr}=K_\mathrm{cr}\Lambda_\mathrm{QCD}$
usually serves as the initial point of the evolution, which is implemented with
the basic variables $Y=\ln(\mu/\mu_0)$ and $\lambda=\ln K_\mathrm{cr}$.
The most frequent choice, $\lambda=0$, corresponds to the
{\it limiting-spectrum} approximation \cite{Azimov:1984np}.
Other recent choices include $\lambda=1.4$ and $\lambda=2.0$
\cite{Perez-Ramos:2013eba}.
Since logarithmic and powerlike corrections become comparable at small values
of $\mu^2$, a judicious choice of $\mu$ is important to prevent strong
correlations.
Motivated by Refs.~\cite{Brodsky:1976mg,Aversa:1990uv,Aguilar:2014tka}, we
choose $\mu^2=R^2Q^2+4M_\mathrm{eff}^2$, where $R$ is the jet radius,
$Q^2=\sqrt{s}$, and $M_\mathrm{eff}$ is the effective gluon mass.
We adopt $R=0.3$ as a typical value from Ref.~\cite{Aversa:1990uv} and
$M_\mathrm{eff}(Q^2)=m^2/[1+(Q^2/M^2)^\gamma]$ with $m=0.375$~GeV,
$M=0.557$~GeV, and $\gamma=1.06$ from Ref.~\cite{Aguilar:2014tka}.

\begin{figure}
\includegraphics[width=\textwidth]{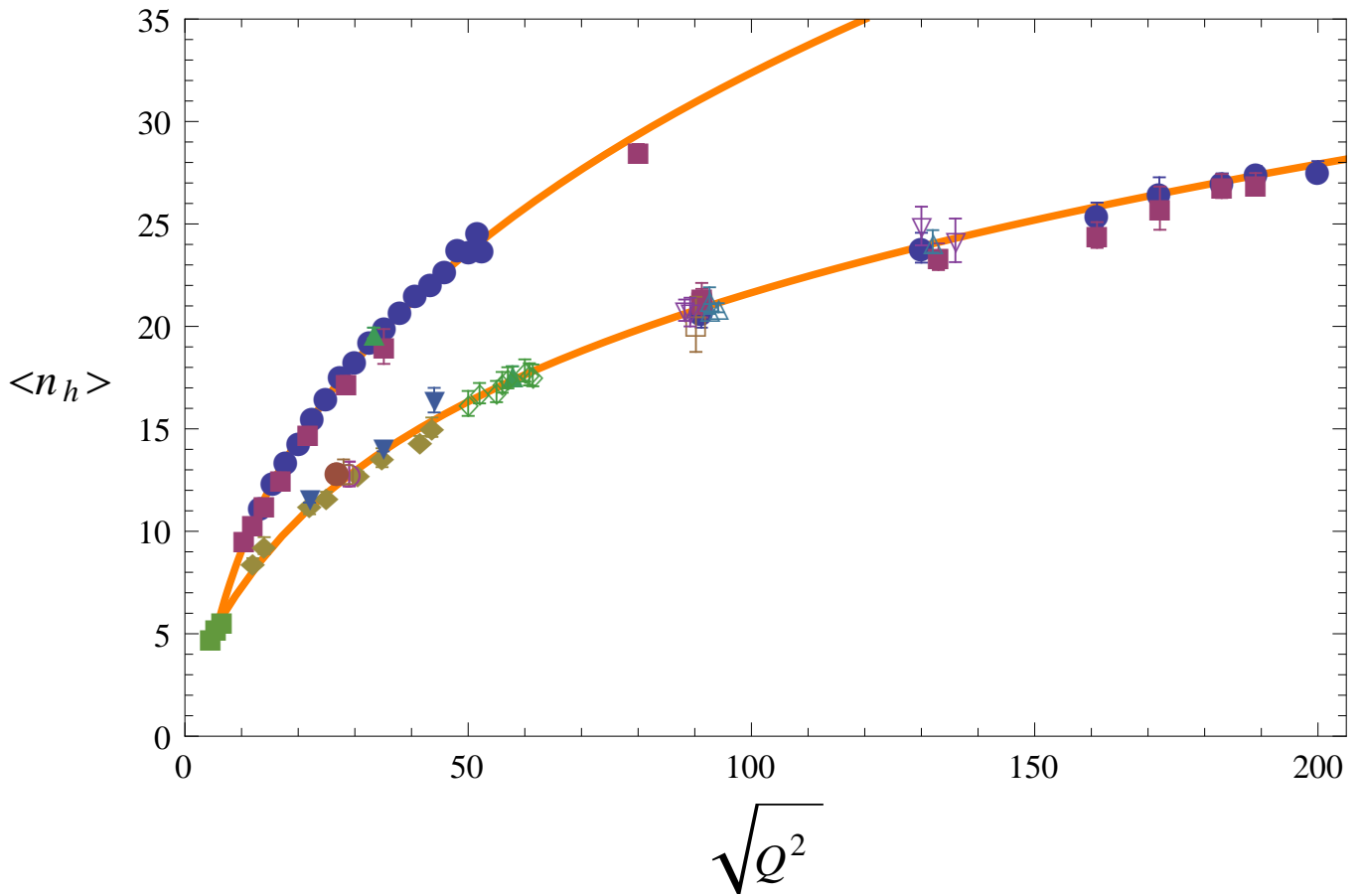}
\caption{Comparison of the experimental data of $\langle n_h(\mu^2)\rangle_q$
(lower curves) and $\langle n_h(\mu^2)\rangle_g$ (upper curves) with the
$\mathrm{NNNLO}_\mathrm{approx}+\mathrm{NNLL}$ fit to them.}
\label{fig:n}
\end{figure}

We are now in a position to perform a global fit to the available measurements
of $\langle n_h\rangle_q$ and $\langle n_h\rangle_g$ for changed hadrons $h$ in
$e^+e^-$ annihilation, which were carefully compiled in
Ref.~\cite{Bolzoni:2013rsa}.
They include 58 and 35 data points, respectively, and 
come from CLASSE CESR with $\sqrt{s}=10$~GeV,
SLAC PEP with 29~GeV,
DESY PETRA with 12--47~GeV,
KEK TRISTAN with 50--61~GeV,
SLAC SLC with 91~GeV,
CERN LEP1 with 91~GeV, and
CERN LEP2 with 130--209~GeV.
The jet algorithms adopted in these experimental analyses are mutually
compatible \cite{Abdallah:2005cy}.
As in Ref.~\cite{Bolzoni:2013rsa}, we choose the reference
scale to be $Q_0=50$~GeV, which roughly corresponds to the geometric mean of
the smallest and largest of the occurring $\sqrt{s}$ values, and put $n_f=5$
throughout our analysis.
As may be seen in Fig.~\ref{fig:n}, our
$\mathrm{NNNLO}_\mathrm{approx}+\mathrm{NNLL}$ fit yields an excellent
description of the experimental data included in it, with a $\chi^2$ per degree
of freedom of $\chi_\mathrm{dof}^2=1.32$.
The fit parameters are determined to be
$\langle n_h(Q_0^2)\rangle_q=16.38\pm 0.05$,
$\langle n_h(Q_0^2)\rangle_g=23.87\pm 0.07$,
$K_\mathrm{cr}=7.09\genfrac{}{}{0pt}{}{+1.71}{-1.21}$, and
\begin{equation}
\alpha_s^{(5)}(M_Z^2)=0.1205\genfrac{}{}{0pt}{}{+0.0016}{-0.0020},
\label{eq:as}
\end{equation}
which nicely agrees with the present world average,
$\alpha_s^{(5)}(M_Z^2)=0.1181\pm0.0011$ \cite{Olive:2016xmw}.
Our fit results turn out to be very insensitive to the precise choice of $Q_0$.
The power corrections turn out to be sizeable, with
$\lambda=1.96\genfrac{}{}{0pt}{}{+0.21}{-0.19}$, 
in agreement with Ref.~\cite{Perez-Ramos:2013eba}.

\begin{figure}
\includegraphics[width=\textwidth]{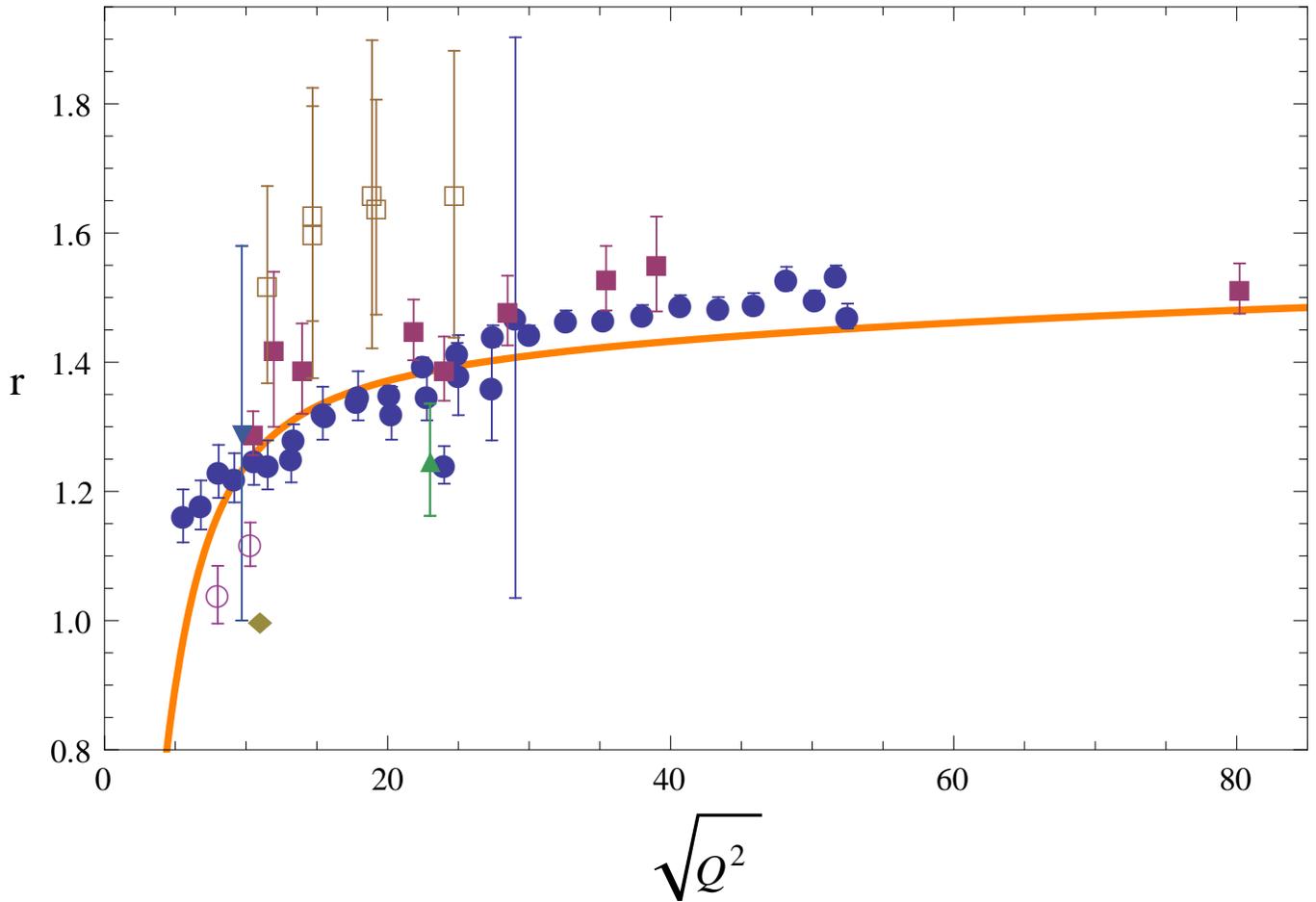}
\caption{Comparison of our $\mathrm{NNNLO}_\mathrm{approx}+\mathrm{NNLL}$
prediction of $r(\mu^2)$ with experimental data excluded from the fit.}
\label{fig:r}
\end{figure}
In Fig.~\ref{fig:r}, we compare our
$\mathrm{NNNLO}_\mathrm{approx}+\mathrm{NNLL}$ prediction for $r$ with the
experimental data compiled in Ref.~\cite{Bolzoni:2013rsa}, which did not enter
our fit.
They were collected at CESR with $\sqrt{s}=10$~GeV,
DESY DORIS~II with 10~GeV,
PEP with 29~GeV,
PETRA with 22--35~GeV,
LEP1 with 91~GeV,
LEP2 with 130--209~GeV, and
FNAL Tevatron with 1.8~TeV.
The agreement is very satisfactory and reassures us of the validity of our
analysis.

In summary, we unraveled an unexpected, SUSY-like relationship between the
NNLL-resummed first Mellin moments of the timelike DGLAP splitting functions
in real QCD, Eq.~(\ref{Basic}), which is $n_f$ independent, and exploited it to
find an exact solution of the DGLAP evolution equation, Eq.~(\ref{apR}),
bypassing the approximate two-step diagonalization procedure used so far in the
literature.
This also allowed us push our knowledge of $r_-$ by one order of $\gamma_0$.
Also incorporating the appropriately transformed $\mathcal{O}(\gamma_0^2)$ and
$\mathcal{O}(\gamma_0^3)$ corrections to $r_+$ as well as power-like
corrections, we performed a global fit to the world data of charged-hadron
multiplicities in quark and gluon jets produced by $e^+e^-$ annihilation and so
extracted the competitive new value of $\alpha_s^{(5)}(M_Z^2)$ in
Eq.~(\ref{eq:as}), which nicely agrees with the present world average.
Our analysis only relies on first principles of QCD and avoids additional
model assumptions, including those inherent to the MLLA.
On top of the physical advantages mentioned above, Eq.~(\ref{Basic}) renders
the otherwise complicated formalism aesthetically pleasing and prompts one to
speculate if there is some unknown higher reason for it.

We thank P.~Bolzoni for collaboration at the initial stage of this research and
O.~L.~Veretin for assistance in the numerical analysis. 
This research was supported in part 
by the German Research Foundation under Grant No.\ KN~365/5-3,
by the National Science Foundation under Grant No.\ NSF~PHY-1125915,
by the Russian Foundation for Basic Research under Grant No.\ 16-02-00790-a,
and by the Heisenberg-Landau Programme.


\begin{thebibliography}{30}

\bibitem{Bjorken:1969ja} 
  J.~D.~Bjorken and E.~A.~Paschos,
  Phys.\ Rev.\  {\bf 185}, 1975 (1969).

\bibitem{Azimov:1984np} 
  Ya.~I.~Azimov, Yu.~L.~Dokshitzer, V.~A.~Khoze, and S.~I.~Troyan,
  Z.\ Phys.\ C {\bf 27}, 65 (1985).

\bibitem{Gribov:1972ri} 
  V.~N.~Gribov and L.~N.~Lipatov,
Yad.\ Fiz.\  {\bf 15}, 781 (1972)
[Sov.\ J.\ Nucl.\ Phys.\  {\bf 15}, 438 (1972)];
  G.~Altarelli and G.~Parisi,
  Nucl.\ Phys.\ {\bf B126}, 298 (1977).

\bibitem{Dokshitzer:1977sg} 
  Yu.~L.~Dokshitzer,
Zh.\ Eksp.\ Teor.\ Fiz.\  {\bf 73}, 1216 (1977)
[Sov.\ Phys.\ JETP {\bf 46}, 641 (1977)].

\bibitem{Almasy:2011eq} 
  A.~A.~Almasy, S.~Moch, and A.~Vogt,
  Nucl.\ Phys.\ {\bf B854}, 133 (2012).

\bibitem{Kniehl:2000fe}
  B.~A.~Kniehl, G.~Kramer, and B.~P\"otter,
  Nucl.\ Phys.\ B {\bf 582}, 514 (2000);
  Phys.\ Rev.\ Lett.\  {\bf 85}, 5288 (2000);
  S.~Albino, B.~A.~Kniehl, and G.~Kramer,
  Nucl.\ Phys.\ {\bf B725}, 181 (2005).

\bibitem{Olive:2016xmw} 
  C.~Patrignani {\it et al.}\ (Particle Data Group),
  Chin.\ Phys.\ C {\bf 40}, 100001 (2016).

\bibitem{Bolzoni:2013rsa} 
  P.~Bolzoni, B.~A.~Kniehl, and A.~V.~Kotikov,
  Nucl.\ Phys.\ {\bf B875}, 18 (2013).

\bibitem{Perez-Ramos:2013eba} 
  R.~P\'erez-Ramos and D.~d'Enterria,
  J. High Energy Phys.\ 08 (2014) 068.

\bibitem{Brodsky:1976mg} 
  S.~J.~Brodsky and J.~F.~Gunion,
  Phys.\ Rev.\ Lett.\  {\bf 37}, 402 (1976);
  K.~Konishi, A.~Ukawa, and G.~Veneziano,
  Phys.\ Lett.\ B {\bf 78}, 243 (1978).

\bibitem{Malaza:1985jd} 
  E.~D.~Malaza and B.~R.~Webber,
  Nucl.\ Phys.\ {\bf B267}, 702 (1986);
  S.~Catani, Yu.~L.~Dokshitzer, F.~Fiorani, and B.~R.~Webber,
  Nucl.\ Phys.\ {\bf B377}, 445 (1992);
  S.~Lupia and W.~Ochs,
  Phys.\ Lett.\ B {\bf 418}, 214 (1998);
  P.~Eden and G.~Gustafson,
  J. High Energy Phys.\ 09 (1998) 015.

\bibitem{Dokshitzer:1991wu} 
  Yu.~L.~Dokshitzer, V.~A.~Khoze, A.~H.~Mueller, and S.~I.~Troian,
  {\it Basics of perturbative QCD},
  Editions Fronti\`eres, Gif-sur-Yvette, 1991.

\bibitem{Mueller:1981ex} 
  A.~H.~Mueller,
  Phys.\ Lett.\ B {\bf 104}, 161 (1981).

\bibitem{Vogt:2011jv} 
  A.~Vogt,
  J. High Energy Phys.\ 10 (2011) 025;
  S.~Albino, P.~Bolzoni, B.~A.~Kniehl, and A.~Kotikov,
  Nucl.\ Phys.\ {\bf B851}, 86 (2011);
  {\bf B855}, 801 (2012).

\bibitem{Kom:2012hd} 
  C.~H.~Kom, A.~Vogt, and K.~Yeats,
  J. High Energy Phys.\ 10 (2012) 033.

\bibitem{Buras:1979yt} 
  A.~J.~Buras,
  Rev.\ Mod.\ Phys.\  {\bf 52}, 199 (1980).

\bibitem{Ellis:1993rb}
  R.~K.~Ellis, Z.~Kunszt, and E.~M.~Levin,
  Nucl.\ Phys.\ {\bf B420}, 517 (1994);
  {\bf B433}, 498(E) (1995);
  A.~Vogt,
  Comput.\ Phys.\ Commun.\  {\bf 170}, 65 (2005).

\bibitem{Kounnas:1982de} 
  C.~Kounnas and D.~A.~Ross,
  Nucl.\ Phys.\ {\bf B214}, 317 (1983);
  A.~P.~Bukhvostov, G.~V.~Frolov, L.~N.~Lipatov, and E.~A.~Kuraev,
  Nucl.\ Phys.\ {\bf B258}, 601 (1985).

\bibitem{Kotikov:1998qt} 
  A.~V.~Kotikov and G.~Parente,
  Nucl.\ Phys.\ {\bf B549}, 242 (1999);
  G.~Cveti\v{c}, A.~Yu.~Illarionov, B.~A.~Kniehl, and A.~V.~Kotikov,
  Phys.\ Lett.\ B {\bf 679}, 350 (2009).

\bibitem{Mueller:1983cq} 
  A.~H.~Mueller,
  Nucl.\ Phys.\ {\bf B241}, 141 (1984);
  J.~B.~Gaffney and A.~H.~Mueller,
  Nucl.\ Phys.\ {\bf B250}, 109 (1985);
  E.~D.~Malaza and B.~R.~Webber,
  Phys.\ Lett.\ B {\bf 149}, 501 (1984).

\bibitem{Dremin:1993vq} 
  I.~M.~Dremin and V.~A.~Nechitailo,
Pis'ma Zh.\ Eksp.\ Teor.\ Fiz.\ {\bf 58}, 945 (1993)
  [JETP Lett.\  {\bf 58}, 881 (1993)];

\bibitem{Capella:1999ms} 
  A.~Capella, I.~M.~Dremin, J.~W.~Gary, V.~A.~Nechitailo, and J.~Tran Thanh Van,
  Phys.\ Rev.\ D {\bf 61}, 074009 (2000).

\bibitem{Dokshitzer:1992iy}
  Y.~L.~Dokshitzer and M.~Olsson,
  Nucl.\ Phys.\ B {\bf 396} (1993) 137.

\bibitem{Aversa:1990uv} 
  F.~Aversa, M.~Greco, P.~Chiappetta, and J.~Ph.~Guillet,
  Phys.\ Rev.\ Lett.\  {\bf 65}, 401 (1990);
  F.~Aversa, P.~Chiappetta, L.~Gonzales, M.~Greco, and J.~Ph.~Guillet,
  Z.\ Phys.\ C {\bf 49}, 459 (1991);
  J.-P.~Guillet,
  Z.\ Phys.\ C {\bf 51}, 587 (1991).

\bibitem{Aguilar:2014tka} 
  A.~C.~Aguilar, D.~Binosi, and J.~Papavassiliou,
  Phys.\ Rev.\ D {\bf 89}, 085032 (2014);
%
  A.~Deur, S.~J.~Brodsky, and G.~F.~de~T\'eramond,
  Prog.\ Part.\ Nucl.\ Phys.\  {\bf 90}, 1 (2016).

\bibitem{Abdallah:2005cy} 
  J.~Abdallah {\it et al.}\  (DELPHI Collaboration),
  Eur.\ Phys.\ J.\ C {\bf 44}, 311 (2005).

\end{thebibliography}
\end{document}